\documentclass[sigconf]{acmart}

\AtBeginDocument{%
  \providecommand\BibTeX{{%
    \normalfont B\kern-0.5em{\scshape i\kern-0.25em b}\kern-0.8em\TeX}}}

\setcopyright{acmcopyright}
\copyrightyear{2021}
\acmYear{2021}
\acmConference[WWW '21]{Proceedings of the Web Conference 2021}{April 19--23, 2021}{Ljubljana, Slovenia}
\acmBooktitle{Proceedings of the Web Conference 2021 (WWW '21), April 19--23, 2021, Ljubljana, Slovenia}
\acmPrice{}

\usepackage{graphicx}
\usepackage{subfigure}
\usepackage{booktabs}
\usepackage{multirow}
\usepackage[ruled]{algorithm2e}
\usepackage{array}
\usepackage{bm}
\theoremstyle{definition}
\newtheorem{definition}{Definition}

\usepackage[flushleft]{threeparttable}

\begin{document}

\title{Automated Creative Optimization for E-Commerce Advertising}

\author{Jin Chen}
\authornote{This work was done when the authors Jin Chen and Gangwei Jiang were at Alibaba Group for intern.}
\affiliation{University of Electronic Science and Technology of China}
\email{chenjin@std.uestc.edu.cn}

\author{Ju Xu}
\affiliation{Alibaba Group}
\email{xuju.xj@alibaba-inc.com}

\author{Gangwei Jiang}
\authornotemark[1]
\affiliation{University of Science and Technology of China}
\email{gwjiang@mail.ustc.edu.cn}

\author{Tiezheng Ge, Zhiqiang Zhang}
\affiliation{Alibaba Group\\
\{tiezheng.gtz,zhang.zhiqiang\}@alibaba-inc.com}

\author{Defu Lian}
\authornote{Corresponding author}
\affiliation{University of Science and Technology of China}
\email{liandefu@ustc.edu.cn}

\author{Kai Zheng}
\affiliation{University of Electronic Science and Technology of China}
\email{zhengkai@uestc.edu.cn}

\renewcommand{\shortauthors}{Jin Chen, Ju Xu, Gangwei Jiang, et al.}

\begin{abstract}
	Advertising creatives are ubiquitous in E-commerce advertisements and aesthetic creatives may improve the click-through rate (CTR) of the products. Nowadays smart advertisement platforms provide the function of compositing creatives based on source materials provided by advertisers. Since a great number of creatives can be generated, it is difficult to accurately predict their CTR given a limited amount of feedback. Factorization machine (FM), which models inner product interaction between features, can be applied for the CTR prediction of creatives. However, interactions between creative elements may be more complex than the inner product, and the FM-estimated CTR may be of high variance due to limited feedback. To address these two issues, we propose an Automated Creative Optimization (AutoCO) framework to model complex interaction between creative elements and to balance between exploration and exploitation. Specifically, motivated by AutoML, we propose one-shot search algorithms for searching effective interaction functions between elements. We then develop stochastic variational inference to estimate the posterior distribution of parameters based on the reparameterization trick, and apply Thompson Sampling for efficiently exploring potentially better creatives. We evaluate the proposed method with both a synthetic dataset and two public datasets. The experimental results show our method can outperform competing baselines with respect to cumulative regret. The online A/B test shows our method leads to a 7\% increase in CTR compared to the baseline.
\end{abstract}


\maketitle

\section{Introduction}
\begin{figure}[t]
	\centering
	\includegraphics[width=0.95\columnwidth]{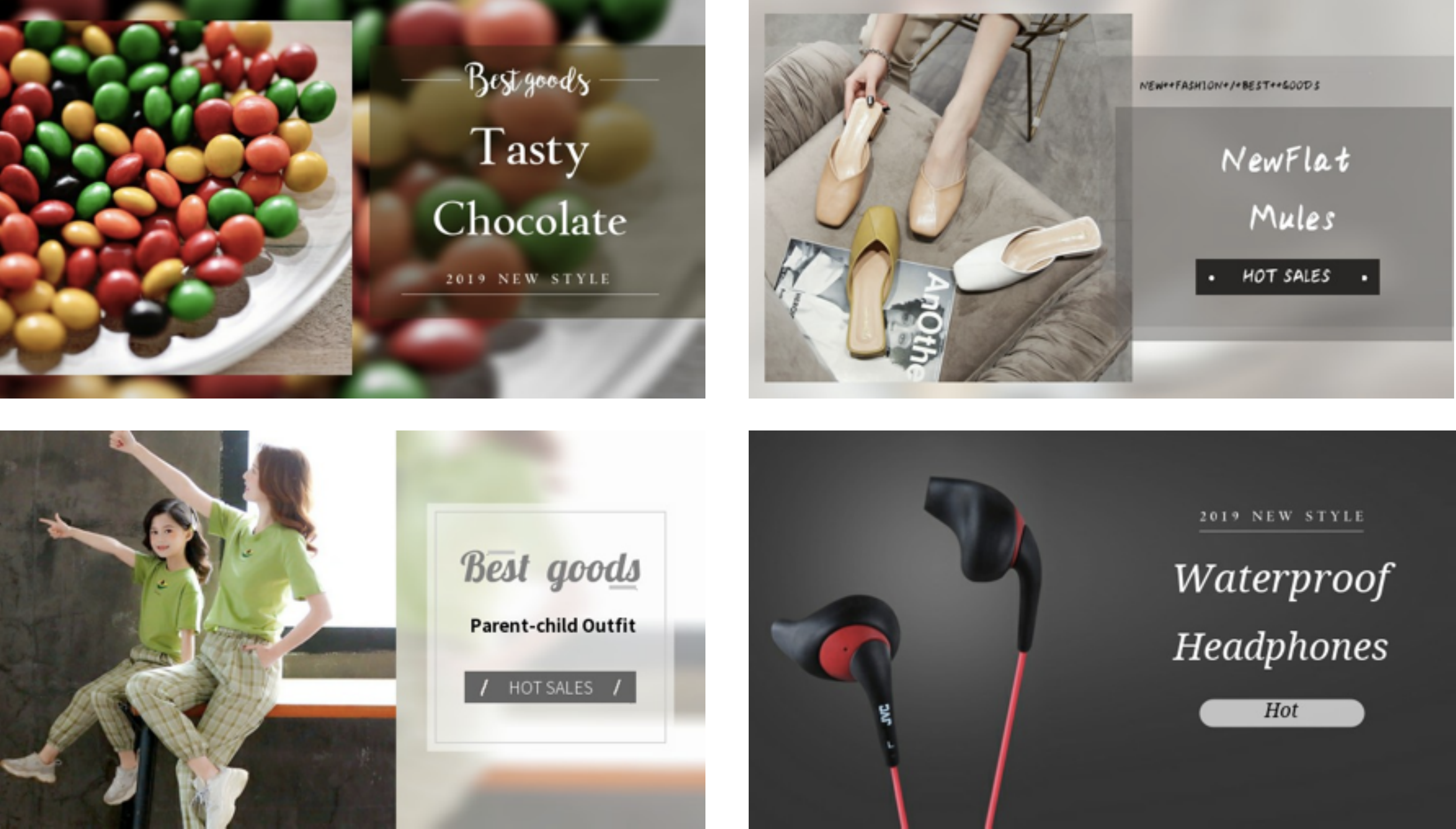}
	\vspace{-0.5em}
	\caption{Advertising Creatives with various elements. These creatives are composited with different templates, product images, textual information, text fonts and different background settings.}
	\label{banners}
	\vspace{-1.0em}
\end{figure}

Online advertisements are ubiquitous in nowadays life, creating considerable revenue for many e-commerce companies. As a common medium of advertisements, advertising creatives, as shown in Fig.~\ref{banners}, can deliver rich product information quickly to users in a visual manner. Appealing creatives improve visual experience and may lead to an increase of click-through rate (CTR), as evidenced by~\cite{azimi2012impact,cheng2012multimedia}. For merchants and e-commerce companies, the increase of CTR can be considered as an indicator of an increase of revenue. Therefore, much attention has been paid to creative design for improving the visual experience. 

Traditionally, advertisers have to employ expert designers to produce attractive creatives and then submit them to platforms as complete images. Each time new products are announced or old products are updated, many creatives of different sizes and styles are required to design and submit to advertising platforms. This leads to an expensive cost for many advertisers. To reduce the cost of repetitive but professional design for advertisers, several high-tech companies set up intelligent advertisement platforms~\cite{hua2018challenges}, which provide instant production services for advertising creatives and remarkably reduces heavy burdens for advertisers. Advertisers only need to provide basic materials to platforms, such as product pictures and textual information. Based on these source materials, the production system produces advertising creatives automatically by compositing arbitrarily designated elements, such as templates, colors of text and sizes of pictures.

In order to ensure the quality of generated creatives, on one hand, they should satisfy basic visual constraints, but this is not the focus of this paper. On the other hand, they should be clicked with high probabilities (i.e., click-through rate) when they are advertised. Intrinsically speaking, the latter corresponds to an optimal selection problem, which faces the following challenges. First, the combinatorial composition of elements leads to an exponential explosion in the number of candidate creatives. Second, because of the limited advertising budget, each product is usually displayed several times within a day. When apportioned to a large number of its generated creatives, the feedback becomes extremely sparse. Furthermore, creatives in E-commerce change frequently over time, so that cumulative feedback for out-of-date products may be not useful any longer. Usually, there are more than 4 million new creatives in a day in a popular advertisement position according to our statistics. Therefore, it is extremely difficult to estimate the click-through rate for each generated creative accurately.

It is possible to apply factorization machines (FM)~\cite{rendle2010factorization} for predicting the  click-through rate (CTR) of each creative. FM models interaction between elements of creative based on inner product, so that creatives with similar composited elements are similarly represented. As a consequence, FM can alleviate the sparsity issue to some extent. However, interactions between elements of creatives may be much more complex than the inner product. For example, we empirically observe that the inner product does not work best for modeling interactions between elements. Moreover, the estimated CTR for each creative may be of high variance due to the extremely sparse feedback. Greedy creative advertising with maximal predicted CTR is usually suboptimal, so that it is essential to efficiently explore potentially better creatives by simultaneously exploiting CTR prediction and uncertainty.

To address these two issues, we propose an Automated Creative Optimization (AutoCO) framework to model complex interaction between elements of creatives and to strike a balance between exploration and exploitation. Particularly, inspired by Automated Machine Learning (AutoML)~\cite{hutter2019automated}, we propose one-shot search algorithms for searching effective interaction functions between elements of creatives efficiently. The interaction function family to search can be defined by extending the multiply operator in the inner product to the operator set \{max, min, plus, multiply, concat\} over operation-aware embedding, and replacing the sum pooling operation with a fully connected network. Following that, following the reparameterization trick in VAE~\cite{kingma2014auto}, we develop stochastic variational inference to estimate the posterior distribution of parameters. Armed with the parameter posterior distribution, we can apply Thompson Sampling~\cite{blundell2015weight} for efficiently exploring potentially better creatives.

The contributions in this paper are summarized as follows.
\begin{itemize}
	\item We propose an Automated Creative Optimization (AutoCO) framework for optimal selection of composited E-commerce creatives. The framework simultaneously models complex interaction between creative elements and strikes a balance between exploration and exploitation, successfully addressing the sparsity issues of feedback.
	\item We empirically observe that the inner product is suboptimal for modeling interaction between elements, based on which we propose a one-shot search algorithm for searching effective interaction functions efficiently.
	\item Based on the reparameterization trick, we develop stochastic variational inference to estimate the posterior distribution of parameters, making it amenable to Thompson Sampling for efficiently exploring potentially better creatives. 
	\item The experiments on both a synthetic dataset and two public datasets show that our method performs better than competing baselines in terms of the cumulative reward and regret, indicating the effectiveness of complex interaction modeling. The online A/B test shows that our method leads to a 7\% increase in CTR, confirming the superiority of our method to baselines.
\end{itemize}

\section{Related Work}
\subsection{Similar Tasks}
With the rapid development of the Internet, the recommendation systems have been proposed to solve the problem of information overload, such as online advertising~\cite{zhou2018deep}, point of interest recommendation~\cite{lian2020geography} and so on. The Creatives are ubiquitous for online advertisements. Some works have paid attention to CTR prediction on ad creatives~\cite{chen2016deep,mo2015image,liu2020category,yang2019learning,zhao2019you} via extracting expressive visual features to increase the CTR. However, there are few works about the optimization for advertising creatives given limited feedbacks. In classic recommendation systems, the negative samplers~\cite{lian2020personalized} are utilized to select informative samplers for solving the data sparsity and the product quantization methods~\cite{lian2020lightrec} have been used for lightweight recommendation.

For online advertising and recommendation, several similar tasks have been studied. LinUCB~\cite{li2010contextual} achieved great success in personalized news recommendations where an efficient contextual bandit is applied. Whole-page optimization, aiming to select the best layout of a home page, is usually defined as a bandit problem~\cite{wang2016beyond}. An efficient multivariate bandit algorithm~\cite{hill2017efficient} is proposed to solve the problem and shows superior performance in real online situations. These similar tasks exploit bandit algorithms to solve the exploration and exploitation dilemma.

\subsection{Success of AutoML for recommendation}
Some success of automated machine learning (AutoML) has been achieved in recommendation systems. Different from regular tasks for AutoML, such as classification in computer vision, search space becomes diverse rather than the neural architectures~\cite{zhong2018practical,zoph2016neural,zoph2018learning}. Searching embeddings of the input with categorical features is exploited for large scale recommendations~\cite{joglekar2020neural}. Another attention is that the search for high order features helps improve the CTR prediction~\cite{10.1145/3394486.3403314}. Furthermore, the search for interaction functions for collaborative filtering reveals the improvements caused by searching operation functions between different user embeddings and item embeddings~\cite{yao2020efficient}.

\section{Preliminaries}
Before we introduce the proposal in detail, we give a brief description of the investigated problem and introduce the classical method. Finally, we provide an overview of our solution.

In this work, only composited creatives are investigated instead of the whole images designed by professional designers. At the time of impression, a product sends a request to the advertisement platform, and the platform instantly selects the creative from candidates for display. There are $M$ elements to be composited and the $i$-th element is a collection of $n_i$ alternatives.

The investigated task in this work is to recommend the optimal composited creative for each product at the time of impression, aiming to increase the overall CTR given limited impressions. Noticing that the generation of the creatives under aesthetic metrics, such as~\cite{li2018layoutgan,vempati2019enabling} is not the focus of this paper.

\subsection{Classical Method}
Typically, each candidate creative should be delivered to the advertising platforms and the user feedback is collected for CTR prediction with point-wise~\cite{chen2016deep}, pair-wise~\cite{he2016vbpr} methods. The creative with a maximal value of predicted CTR is then selected to be displayed to increase the overall click-through rate. But the multiple elements are composited combinatorially, resulting in an exponential explosion in the number of creatives. For example, given 4 templates, 10 fonts, 10 colors, and 5 picture sizes, 2,000 creatives can be composited for one of the product images. The collection of sufficient samples for model training is time-consuming and expensive. Thus there is a trade-off between exploration and exploitation.

\textbf{Multi-Armed Bandit} is usually a classical option for creative optimization where the composited creatives are compared to bandits. These algorithms generally include the following processes. At a discrete-time $t\in [1,...,T]$, a creative $c$ is selected and revealed with a reward $R_{c_t}\in \{ 0,1\}$. An impression is a scenario when an ad is shown to a user in the advertising industry and $R=1$ represents the user clicked the selected creative. Within the limited impressions, the multi-armed bandit aims to minimize the cumulative regret $\sum_{t=1}^{T} \left( R_{c^*} - R_{c_t} \right)$, where ${c^* = \arg \max \mathbb{E}(R_c)}$ represents the optimal creative with the maximal expected reward. In this work, the expected reward corresponds to the value of CTR. 

The composited creatives sometimes share common elements, so that making connections can help estimate the expected reward of various creatives. The linear function to model the connections has been successfully applied in this situation~\cite{chu2011contextual}, where the expected reward of the creative is assumed as a linear function with respect to the feature. The expected reward for each creative is additionally measured by a upper confidence bound. The linear models ensure the theoretical lower bound for regret. However, the expressiveness of linear models is limited. Such models can not efficiently capture the commonality given significantly large search space. This limitation motivates us to design a more effective algorithm under the huge creative space given extremely sparse feedback.

\subsection{Solutions Overview}
There are two key components in a bandit problem: (1) Assumption of the expected reward (2) Exploration methodology. For these two parts, we design effective algorithms to improve the performance in terms of overall CTR under the huge creative space. The proposed framework consists of the following phases:
\begin{enumerate}
	\item \textbf{CTR Estimation}: We focus on the interactions between different elements to leverage commonality between numerous creatives, and search for the interaction functions to capture suitable representations depending on AutoML methods.
	\item \textbf{Efficient Exploration}: To reduce the high variance caused by sparse feedback, Thompson Sampling, a classical and efficient exploration method, is exploited. The posterior approximation under complex interaction functions is estimated through variational inference in an end-to-end manner.
\end{enumerate}

\section{Interaction Function Search for CTR Estimation}
\subsection{Interaction Function Analysis}
In a similar task, the optimal selection of layouts for web pages~\cite{hill2017efficient}, the interactions between different elements are captured as the way in the Poly2~\cite{chang2010training} model. But the parameter space becomes seriously large with the increasing number of features. One solution is the factorization machines (FMs)~\cite{rendle2010factorization}, which utilize the inner product as the interaction functions and achieve success in CTR prediction~\cite{guo2017deepfm}. The inner product of the pairwise feature interactions is as:
\begin{displaymath}
<\bm{e}_i,\bm{e}_j> = \sum_{k=1}^{d} e_{ik} \cdot e_{jk} 
\end{displaymath}
where $\bm{e}_i\in \mathbb{R}^{d}$ is the low-dimension embedding vector for the $i$-th feature. $\bm{e}_i=V_i \bm{x}_i$. $V_i\in \mathbb{R}^{d \times l_i}$ is the embedding matrix with respect to the $i$-th feature field and $\bm{x}_i\in \mathbb{R}^{l_i}$ denotes the one-hot encoding vector of the $i$-th feature. Multi-hot vectors or continuous value vectors are also available. $d$ is the embedding dimension and $l_i$ represents the size of feature space.

However, the inner product may not yield the best performance in many recommendation tasks due to the complex nature of interactions. Inspired by the recent work SIF~\cite{yao2020efficient}, which searches for the interaction functions between the user and item embeddings, we are encouraged to capture the complicated interactions between different features. For the sake of simplicity, we state the problem of searching for the interaction functions between different feature fields in this investigated task.
\begin{definition}[Interaction Functions Search]
	Find an interaction function $o_{ij}\in \mathcal{O}$ for each pair of embedding vectors $(\bm{e}_i, \bm{e}_j)$. The object is to minimize the loss over the observed data. 
\end{definition}

$\mathcal{O}$ is the operation set containing several interaction functions. We utilize $|\mathcal{O}|$ to represent the number of operations in $\mathcal{O}. $There are $L$ embedding vectors and the number of the interaction functions to be selected is $K = \mathrm{C}_{L}^2$. Thus the time complexity for searching the optimal interaction function is $O(|\mathcal{O}|^K)$. In the following chapters, we will introduce the search space and the search algorithm respectively.

\subsection{Search Space}
\begin{table*}[ht]
	\caption{AUC and the Relative improvements on different interaction functions compared with FM}
	\label{offline_different_func_auc}
	\large
	\begin{tabular}{l|c|c|c|c|c|c|c|c}
		\toprule
		& Multiply & Concat & Plus & Max & Min & LR & IFS(Alg.\ref{one_shot_interaction_search}) & FM \\
		\midrule
		Average AUC&0.6039&0.6044&0.6037&0.6039&0.6023&0.5285&0.6050&0.5947\\
		Standard Error&0.0017&0.0004&0.0007&0.0013&0.0008&0.0019&0.0003&0.0004\\
		Relative Improvements& 1.54\% & 1.62\% & 1.50\% & 1.53\% & 1.27\% & -11.13\% & 1.71\% & - \\ 
		\bottomrule
	\end{tabular}
\end{table*}

We pick the following simple and popular operations as mentioned in the previous work~\cite{yao2020efficient}. (1) \textit{Concat}: $[\bm{p};\bm{q}]$, (2) \textit{Multiply}: $\bm{p} \circ \bm{q}$, (3) \textit{Plus}: $\bm{p}+\bm{q}$, (4) \textit{Max}: $\max(\bm{p},\bm{q})$, (5) \textit{Min} : $\min(\bm{p},\bm{q})$. The concat is a vector-wise operation and the other four operations are element-wise. Similar with SIF, we exploit a full connected layer for each function. The FC controls the output size of different operations to be consistent.

Before we focus on the approach of the efficient search for interaction functions, several offline experiments are conducted to demonstrate that different interaction functions between feature fields yield different performances. The baseline approaches involve the factorization machine and the logistics regression without interactions.

As shown in Table \ref{offline_different_func_auc}, different operations have been studied on the collection of about one thousand products and their 300 thousand composited creatives. Nearly three million pieces of data are collected. We use the regular CTR prediction with the binary cross-entropy loss for optimization. The five selected operation functions have better performances than the FM models and have different degrees of improvement. The result provides evidence that there exist more approximate interaction functions between the different elements.

Considering all the interactions between different embedding vectors, the search space is up to $5^{K}$ and is much more complicated. The various operators between different embedding vectors may enhance the expressiveness of the CTR predictor and then lead to better performances than the FM-based models.

\subsection{Efficient Search Strategy}
A straightforward idea to obtain the optimal interactions between different feature fields is to traverse all the combinations. However, the time complexity of collecting the best interactions is $O(|\mathcal{O}|^K)$, which is NP-hard and unacceptable in our situation. To deal with the issue, we apply an efficient search algorithm inspired by previous AutoML algorithms.

\begin{figure*}[ht]
	\centering
	\includegraphics[width=0.99\textwidth]{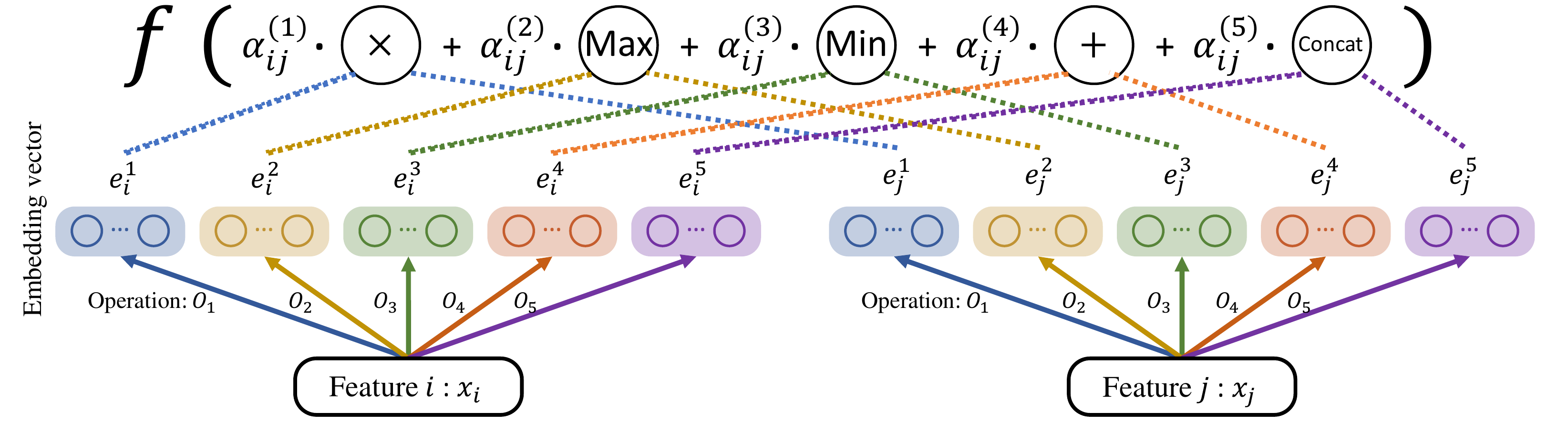}
	\vspace{-0.4em}
	\caption{Representation of the search space with the Operation-Aware Embedding module.}
	\label{ofm_framework}
\end{figure*}
\subsubsection{One-Shot Search Algorithm}
Neural architecture search is firstly proposed in~\cite{zoph2017neural}, where reinforcement learning is employed for searching architectures. After then, more search-based algorithms have been proposed (e.g., evolution algorithm~\cite{liu2018hierarchical}, greedy search~\cite{liu2018progressive}). However, these algorithms cost much search time due to the huge and discrete search space. Recently, the one-shot architecture search algorithms achieve huge success depending on the continuous search space and greatly shorten the time, such as DARTS~\cite{liu2018darts}, SNAS~\cite{xie2018snas}, NASP~\cite{yao2019efficient}. These algorithms jointly optimize the weights for different operators and model parameters by stochastic gradient descent.

Following SIF~\cite{yao2020efficient}, we relax the choices among operations as a sparse vector in a continuous space. For each pair of feature fields $i$ and $j$, $\bm{\alpha}_{ij} \in  \mathbb{R}^{|\mathcal{O}|}$ denotes the weight vector and the $k$-th element $\alpha^{(k)}_{ij}$ corresponds to the weight of the $k$-th operator $\mathcal{O}_{ij}^{(k)}$. Now the operation $\bar{\mathcal{O}}_{ij}$ between the feature pair $i$ and $j$ is formulated as follows:
\begin{displaymath}
\begin{aligned}
& \bar{\mathcal{O}}_{ij}(\bm{X}) = \sum_{k=1}^{|\mathcal{O}|} \alpha^{(k)}_{ij} \mathcal{O}^{(k)}_{ij} (\bm{X}), {\text{  where }} \bm{\alpha}_{ij} \in \mathcal{C}_1 \cap \mathcal{C}_2.\\
& \mathcal{C}_1 = \{ \bm{\alpha}| \parallel \bm{\alpha} \parallel_0 = 1  \}, \qquad \mathcal{C}_2 = \{\bm{\alpha} | 0 < \alpha^{(k)} < 1 \}
\end{aligned}
\end{displaymath} 
The constraint $\mathcal{C}_1$ ensures that only one operator would be selected, and the constraint $\mathcal{C}_2$ makes the algorithm more stable. Thus, the search problem is defined as:
\begin{equation}
\label{nas_objective}
\begin{aligned}
\min_{\bm{\Theta},\bm{\alpha}} \quad &F(\bm{\Theta},\bm{\alpha})  = \sum_{ \left(\bm{x},y \right) \in \mathcal{D}}  \ell \left(\mathcal{I}[\bm{x},\bm{\alpha},\bm{\Theta}],y \right) \\
{\text{s.t.  }} \quad &\bm{\alpha}  =  \{ \bm{\alpha}_{ij}|i=1,...,L,j=i+1,...,L\}\\
&\bm{\alpha}_{ij} \in \mathcal{C}_1 \cap \mathcal{C}_2\\   
\end{aligned}
\end{equation}
where $    \mathcal{I}[\bm{x}, \bm{\alpha}, \bm{\Theta}]  = \sum_{i}^{L} \sum_{j=i+1}^L \sum_{k=1}^{|\mathcal{O}|} \alpha^{(k)}_{ij}\mathcal{O}_{ij}^{(k)}(\bm{e}_i,\bm{e}_j) $, $\bm{\alpha}$ denotes the operation weights for all the $|\mathcal{O}|^{K}$ interactions. $\ell$ is the binary cross-entropy loss in this work. $\bm{\Theta}$ represents the parameters for embedding matrices and the fully connected layers. For the sake of simplicity, we temporarily omit the expression of the first-order term and the bias in FMs models. Following previous successful work with proximal steps~\cite{yao2020efficient,yao2019efficient}, we utilize the same optimization method for efficient searching.

\begin{algorithm}[t]
	\caption{Interaction Function Search (IFS)}\label{one_shot_interaction_search}
	\LinesNumbered
	\KwIn{Observations $\mathcal{D}$, learning rate $\eta$}
	\While(){Not Converge}{
		Select the interaction function $\bar{\bm{\alpha}}_{ij} = prox_{\mathcal{C}_1}(\bm{\alpha}_{ij})$ for each pair\;
		Calculate $F({\bm{\Theta}},\bar{\bm{\alpha}})$ according to Eq (\ref{nas_objective}) and (\ref{nas_ofm})\;
		Update continuous $\bm{\alpha}$ of operation weights
		$\bm{\alpha} = prox_{\mathcal{C}_2}(\bm{\alpha} - \eta \nabla_{\bar{\bm{\alpha}}}F({\bm{\Theta}}, \bar{\bm{\alpha}})$\;
	}
	\KwRet{Interaction functions between each pair of embedding vectors parameterized by $\bm{\alpha}$}
\end{algorithm}

\subsubsection{Operation-aware Embedding}
An interesting point proposed in FFM~\cite{juan2016field} and FwFM~\cite{pan2018field}, that features from one field often interact differently with that from other fields, prompts us the potential distinct influence of operators on feature embeddings. Noticing that the joint optimization for the operation weights and embedding vectors, we design the operation-aware Embedding module, shorted as OAE, which assigns different embeddings for each operator. Fig.~\ref{ofm_framework} represents the interactions of different operations.

According to the operation-aware Embedding, $\mathcal{I}$ is redefined as:
\begin{equation}
\label{nas_ofm}
\mathcal{I}[\bm{x}, \bm{\alpha}, \bm{\Theta}] = \sum_{i=1}^{L} \sum_{j=i+1}^{L} \sum_{k=1}^{|\mathcal{O}|}\alpha_{ij}^{(k)} \mathcal{O}_{ij}^{(k)}(\bm{e}^{(k)}_i,\bm{e}^{(k)}_j)
\end{equation}
The embedding $\bm{e}^{(k)}_{i}$ represents the embedding vector corresponding to the $k$-th operation. Observing the original Eq (\ref{nas_objective}), the shared embedding matrices would obtain gradients of different magnitudes with the change of the operator. A comprehensible example is a difference between the plus and multiply operations. The frequent change would affect the computation of the gradient and result in poor convergence. The introduction of the operation-aware embedding removes the interferences of other operations when optimizing the embeddings and may lead to better learning. The final algorithm for interaction function search with the OAE module is described in Alg. \ref{one_shot_interaction_search}.

After the model training, the interaction function $O_{ij}$ is selected by $O_{ij} = \arg \max \bm{\alpha}_{ij}$. The expected reward of the candidate creative $c$ is followed as:
\begin{equation}
\label{inference}
\mathbb{E}[r|\bm{x}_{c}] = \sigma\left( \sum_i^L \sum_{j=i+1}^L O_{ij}\left(\bm{e}_{i}^{O_{ij}},\bm{e}_{j}^{O_{ij}} \right) \right) 
\end{equation}
where $\sigma(x) =  \frac{1}{1+e^{-x}}$ is the sigmoid function. Then, the creative with the highest expected reward will be selected and revealed with a reward.

\section{Efficient Exploration via Thompson Sampling}
As mentioned before, the platform can composite hundreds of creatives. But the limited impressions of the products result in the extreme sparsity of feedback. Each candidate creative would be displayed with few impressions so that the CTR predictor comes with high variance. Thus, an efficient exploration method is introduced to reduce the high variance. 

\subsection{Variational Inference for Thompson Sampling}
\label{TS_vi}
Thompson Sampling~\cite{thompson1933likelihood} is a popular bandit algorithm to balance exploration and exploitation. The main idea is to assume a simple prior distribution of the reward distribution of each candidate, and at any time $t \in [1,...,T]$, recommend a candidate proportionally to the posterior distribution of being optimal. As declared in~\cite{agrawal2013thompson}, instead of solving the posterior distribution of the reward distribution, Thompson Sampling for contextual bandits samples the model parameters. A general process usually includes the following steps:
\begin{enumerate}
	\item Sample model weights from the posterior $\tilde{\bm{\bm{\Theta}}} \sim p(\bm{\bm{\Theta}} | \mathcal{D})$;
	\item Recommend the creative $c = \arg\max P(r|\bm{x},\tilde{\bm{\bm{\Theta}}})$;
	\item Receive reward $r$;
	\item Update past observations $\mathcal{D}$ and the posterior $p(\bm{\bm{\Theta}} | \mathcal{D})$.
\end{enumerate}

For linear models, given the Gaussian distribution as priors, the posterior has closed-form solutions following Bayesian rules~\cite{agrawal2013thompson,li2012unbiased}. However, by introducing the complex operations between different feature fields, the posterior distribution is intractable to be directly solved. Thanks to the previous success of posterior approximation in neural networks~\cite{blundell2015weight,kingma2014auto}, we can exploit variational inference for approximation to fit in the Thompson Sampling method.

Assuming that the model parameters satisfy the Gaussian distributions, variational learning finds the parameters $\mu,\Sigma$ of the approximation posterior $q(\bm{\bm{\Theta}}; \mu, \Sigma) = \mathcal{N}(\mu, \Sigma)$ by minimizing the Kullback-Leibler (KL) divergence between the approximation posterior and the truth posterior:
\begin{displaymath}
D_{KL}\left( q(\bm{\bm{\Theta}}; \mu, \Sigma) || p(\bm{\bm{\Theta}} | \mathcal{D}) \right) = E_{\mathcal{N}(\bm{\bm{\Theta}}; \mu, \Sigma)} \left[ \log \frac{\mathcal{N}(\bm{\Theta}; \mu, \Sigma)}{p(\bm{\Theta} | \mathcal{D})} \right]
\end{displaymath}
where $\mathcal{D}$ is the set of observed data and each sample is the feature vector $\bm{x}$ and corresponding reward $y$. Following Bayesian rules, 
the KL-divergence can be rewritten as:
\begin{displaymath}
E_{\mathcal{N}(\bm{\Theta}; \mu, \Sigma)} \left[ \log \frac{\mathcal{N}(\bm{\Theta}; \mu, \Sigma)}{p(\bm{\Theta}) p(\mathcal{D}|\bm{\Theta})} \right] + \log p(\mathcal{D})
\end{displaymath}

The component $\log p(\mathcal{D})$ is independent of the learning parameters $\mu$ and $\Sigma$, so that the object equals to minimize the following equation:
\begin{displaymath}
\begin{aligned}
\mathcal{J} & = E_{\mathcal{N}(\bm{\Theta}; \mu, \Sigma)} \left[ \log \mathcal{N}(\bm{\Theta}; \mu, \Sigma) - \log p(\bm{\Theta}) - \log p(\mathcal{D}|\bm{\Theta}) \right] \\
& = D_{KL}\left( \mathcal{N}(\bm{\Theta}; \mu, \Sigma) || p(\bm{\Theta}) \right) - E_{\mathcal{N}(\bm{\Theta}; \mu, \Sigma)} \left[  \log p(\mathcal{D}|\bm{\Theta}) \right]
\end{aligned} 
\end{displaymath}

According to Equation (\ref{inference}), the likelihood can be formulated as:
\begin{displaymath}
p(\mathcal{D}|\bm{\Theta}) = \prod_{(\bm{x},y) \in \mathcal{D}} \left(\sigma\left( \mathcal{I}[\bm{x}] \right) \right)^{y} \left( 1-\sigma(\mathcal{I}[\bm{x}]) \right)^{1-y}
\end{displaymath}

The logarithm can be simplified as:
\begin{displaymath}
\small
\begin{aligned}
\log p(\mathcal{D}|\bm{\Theta}) & = \sum_{(\bm{x},y) \in \mathcal{D}}  y \log \sigma\left( \mathcal{I}[\bm{x}] \right) + (1-y) \log \left(1- \sigma\left( \mathcal{I}[\bm{x}] \right) \right) \\
& = \sum_{(\bm{x},y) \in \mathcal{D}} \ell[\bm{x},y| \bm{\Theta}]
\end{aligned}
\end{displaymath}

Usually, the stochastic variational inference is applied to estimate $E_{\mathcal{N}(\bm{\Theta}; \mu, \Sigma)} \left[  \log p(\mathcal{D}|\bm{\Theta}) \right]$. Suppose that the variational posterior is a diagonal Gaussian distribution, we reparameterize the model parameters which yields a posterior sample of the parameters as $\bm{\Theta}^{(s)} = \mu + \Sigma^{\frac{1}{2}} \circ \epsilon $. The noises $\epsilon \sim \mathcal{N}(0,1)$ has the same size as $\bm{\Theta}$ and $\circ$ is point-wise multiplication. Finally, the objective function is formulated as:
\begin{equation}
\small
\mathcal{L}[\bm{\Theta}^{(s)}; \mu,\Sigma] = D_{KL}\left( \mathcal{N}(\bm{\Theta}; \mu, \Sigma) || p(\bm{\Theta}) \right) - \sum_{(\bm{x},y) \in \mathcal{D}} \ell[\bm{x},y| \bm{\Theta}^{(s)}]
\label{variational_posterior}
\end{equation}
where the prior $p(\bm{\Theta})$ is assumed as Gaussian distributions in this work. The KL divergence of two Gaussian distributions has an analytical expression. Thus, stochastic gradient descent can be obtained to minimize the objective function. At each iteration, we draw same Gaussian noise $\epsilon$ and evaluate Equation (\ref{variational_posterior}) and its derivations w.r.t. $\mu$, $\Sigma$. The approximation posterior is then estimated as the parameter updates within training iterations. 

Integrated with the interaction function search, the whole optimization procedure of Thompson Sampling is declared in Alg. \ref{one_shot_interaction_search_ts}. The proposed approach is updated in an end-to-end manner without multiple repetitions of running after the interaction functions update. As mentioned in~\cite{blundell2015weight,kong2020sdenet}, Non-Bayesian approaches, which trains multiple models for estimating the expected reward, is prohibitively expensive in practice. The application of variational inference facilitates a quick response in a live environment. 

\begin{algorithm}[t]
	\caption{Interaction Function Search with Thompson Sampling}
	\label{one_shot_interaction_search_ts}
	\LinesNumbered
	\KwIn{Observations $\mathcal{D}$, learning rate $\eta$}
	Initialize the parameters $\mu$ and $\Sigma$\;
	\While(){Not Converge}{
		Select the interaction function $\bar{\bm{\alpha}}_{ij} = prox_{\mathcal{C}_1}(\bm{\alpha}_{ij})$ for each pair\;
		Sample noise $\epsilon \sim \mathcal{N}(0,1)$\;
		Reparameterize embedding parameters $\tilde{\bm{\Theta}} = \mu + \Sigma^{\frac{1}{2}} \circ \epsilon$\;
		Calculate $F(\tilde{\bm{\Theta}},\bar{\bm{\alpha}})$ according to Eq (\ref{nas_objective})(\ref{nas_ofm})\;
		Update continuous $\bm{\alpha}$ of operation weights
		$\bm{\alpha} = prox_{\mathcal{C}_2}(\bm{\alpha} - \eta \nabla_{\bar{\bm{\alpha}}}F(\tilde{\bm{\Theta}}, \bar{\bm{\alpha}})$\;
		Update $\mu$, $\Sigma$ of variational distribution of embedding matrices corresponding to the selected operation\;
	}
	\KwRet{Interaction functions parameterized by $\bm{\alpha}$, Model parameters parameterized by $\mu$, $\Sigma$}
\end{algorithm}

\subsection{Automated Creative Optimization for E-Commerce Advertisement}
Following the above Thompson Sampling method, we can sample a new set of model parameters to estimate the expected reward and pick the creative with the highest value for display. The creative is then delivered with a reward. After getting the new observations, we search for the interaction functions and embedding matrices simultaneously, as shown in Alg. \ref{one_shot_interaction_search_ts}. This end-to-end optimization facilitates efficient search and efficient exploration for creative optimization. The final framework of automated creative optimization is shown in Alg. \ref{nas_ts}. The algorithm would converge as the observations accumulate. 

\begin{algorithm}[t]
	\caption{\textbf{Auto}mated \textbf{C}reative \textbf{O}ptimization for E-Commerce Advertisement (AutoCO)}\label{nas_ts}
	\LinesNumbered
	\KwIn{Feature vectors $\bm{X}$ of the products and candidate creatives}
	Initialize $\mu$ and $\Sigma$ randomly for embedding matrices\;
	Initialize the interaction functions parameterized by $\bm{\alpha}$\;
	Observations $\mathcal{D} = \emptyset $ \;
	\For{t=1,2,...,T}{
		Receive the request of product $p$\;
		Select the interaction functions $O_{ij} = \arg\max \bm{\alpha}_{ij}$ for each pair of interactions\;
		Obtain the embedding vectors $\tilde{\bm{e}}$ after sampling parameters $\bm{\Theta} \sim \mathcal{N}(\mu,\Sigma)$\;
		The candidate set $\mathcal{C}$ corresponding to the product $p$\;
		Choose the candidate $c_t = \arg \max_{c\in \mathcal{C}} \mathbb{E}[r|\bm{x}_{p,c}]$ according to Eq (\ref{inference})\;
		Receive a binary reward $R_t=\{0,1\}$\;
		Add the new observation $(\bm{x}_{p,c_t}, R_t )$ into $\mathcal{D}$\;
		Update $\mu$, $\Sigma$, $\bm{\alpha}$ according to Alg. \ref{one_shot_interaction_search_ts}\;
	}
\end{algorithm}

To this end, we perform an automated interaction function search improving the expressiveness of the model and utilize the Thompson Sampling via variational inference for efficient exploration to reduce the high variance caused by limited feedback.

\section{Experiments}
In this section, experiments\footnote{https://github.com/alimama-creative/AutoCO} are conducted on the synthetic dataset and public datasets to answer the following questions:
\begin{itemize}
	\item \textbf{Q1}: How the automated interaction function search improve the performance compared with the FM model?
	\item \textbf{Q2}: Can the proposed Thompson Sampling with variational inference for posterior approximation explore efficiently with limited feedback?
\end{itemize}
To further examine the performance of our proposed approach, we conduct the experiment on a live production system.

\subsection{Synthetic Results}
\subsubsection{\textbf{Simulation Data}}
Following the synthetic experiments in the predictor work~\cite{hill2017efficient}, we produce the synthetic data depending on the assumed CTR predictor and simulate the reward for the proceeding of bandit problems. 
There are five elements for compositing creatives, including template, picture size, text font, background blur, and background color. We generate the expected reward over 167 products which uses the representation in Equation~\ref{inference}. On average, each product has 67 composited creatives satisfying visual constraints. The operators for different embedding vectors are initialized randomly and the parameters of embedding matrices are sampled from Gaussian distributions. Finally, each creative has an expected reward (CTR), which is the success probability in the Bernoulli trails to simulate the user feedback. The average value of the generated CTR is 0.0885. 

At each trial, we sample a product and the algorithm recommends the creative which owns the estimated maximal CTR. Every 10,000 trails are collected for updating and training to simulate the delayed feedback in real production systems. Each algorithm is run multiple times with 20 batches to reduce the bias caused by randomness. The empirical studies are running in a Linux operating system with the Tesla P100 GPU for accelerating.

\subsubsection{\textbf{Baselines}}
We choose the following typical algorithms for comparisons where parameters are tuned with the best performance in tems of the overall cumulative CTR.
\begin{itemize}
	\item \textbf{Uniform}: The creative is selected uniformly at random. 
	\item \textbf{Egreedy}: Recommend a creative at random with probability $0.1$ and recommend the creative with the best statistical reward with probability $0.9$.
	\item \textbf{LinUCB}~\cite{li2010contextual}: Each candidate creative has the corresponding linear models with the observed contextual information. LinUCB recommends the creative according to the estimated reward with a upper confidence bound. 
	\item \textbf{LinTS}: Similarly with LinUCB, the interactions are constructed by linear models. LinTS utilizes Thompson Sampling for exploration depending on the Bayesian linear regression.
	\item \textbf{MVT}~\cite{hill2017efficient}: MVT captures the interactions with a linear model and utilizes Thompson Sampling proposed in this work for exploration.
\end{itemize}

We add another two \textbf{FM}-based algorithms \textbf{FM} and \textbf{FM-TS}. \textbf{FM} recommends the creative which owns the maximal estimated CTR without exploration while \textbf{FM-TS} exploits Thompson Sampling as declared in Section \ref{TS_vi}.

Our proposed method \textbf{AutoCO} automatically searches interaction functions between different feature fields and utilizes the same exploration method as FM-TS. 

\begin{figure}[t]
	\centering
	\includegraphics[width=\columnwidth]{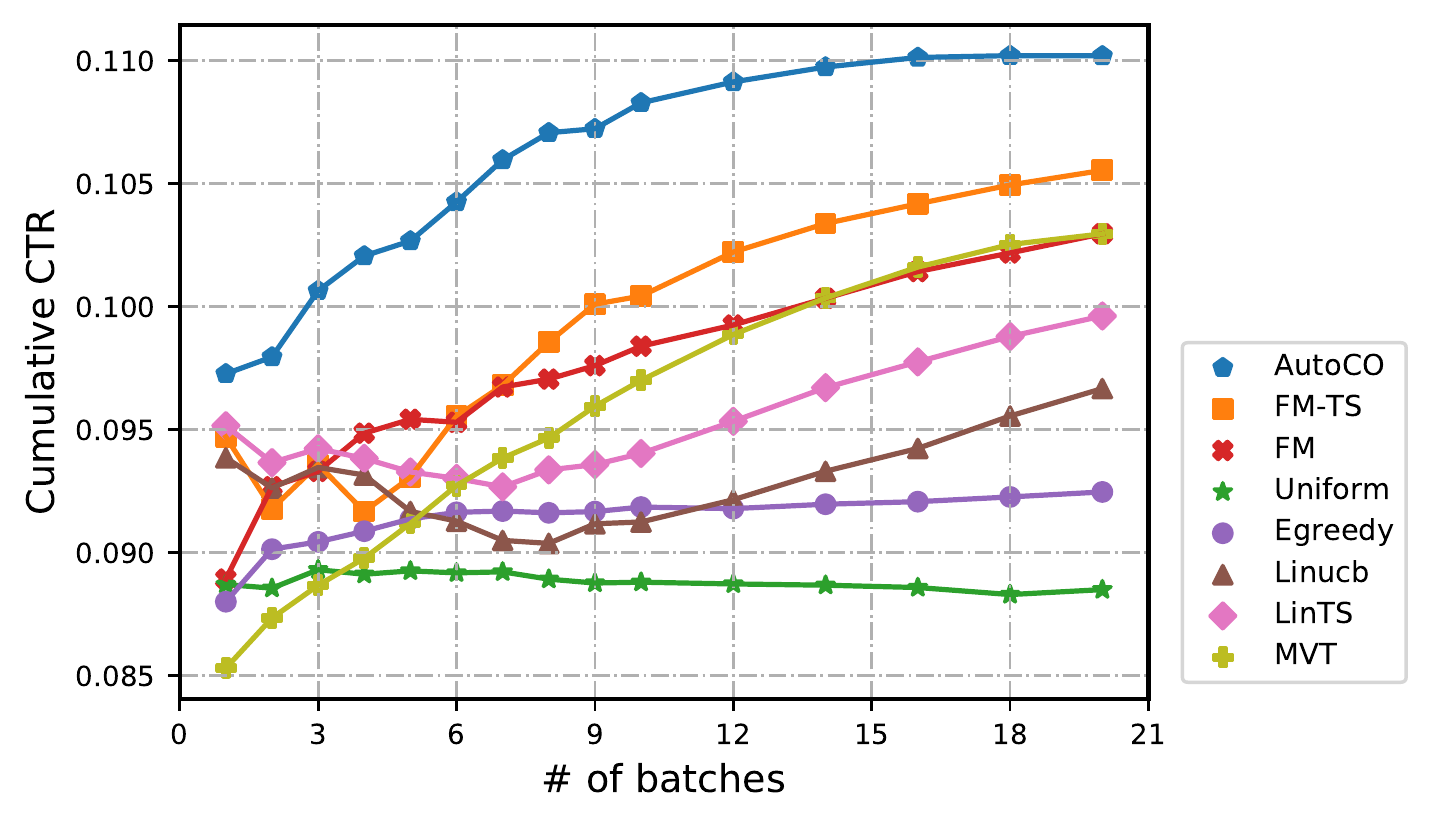}
	\caption{Result of CTR on the simulated data compared with baseline algorithms. Experiments are run for 5 repetitions.}
	\label{baseline_comparisons_nasts}
\end{figure}

\subsubsection{\textbf{Metrics}}
Generally, bandit algorithms are evaluated in terms of cumulative regret~\cite{sutton2018reinforcement} which represents the cost before finding the optimum. In practice, the regret is the difference between the expected reward of the optimal strategy and the empirical performance of the conducted algorithm. The lower regret is better. The value of the cumulative reward is complementary to the cumulative regret and higher is better. We also report the normalized cumulative reward, which is actually the CTR in online advertisements.

\subsubsection{\textbf{Learning Effectiveness}} We perform 20 batches, i.e.2 million impressions to test each algorithm. We report the average performance of 5 repetitions experiments in Fig.~\ref{baseline_comparisons_nasts}. There are several findings in our empirical studies.

\textit{Finding 1}: The interactions between different elements enrich the commonality leverage and lead to better performances for CTR prediction. The linear models (e.g., LinUCB and LinTS) which only build the linear relationship between elements converge to worse creatives although they have comparable performances at the beginning. The expressive representations for the interactions help model the commonality of numerous creatives and improve the performances of the overall CTR.

\textit{Finding 2}: The proposed AutoCO method shows superior performances during the whole process. The significant improvement comes from two key points: (1) Automated interaction function search; (2) Efficient exploration with variation based Thompson Sampling. To investigate the different influences of these two parts, we conduct two experiments. 

\begin{figure}[t]
	\centering
	\includegraphics[width=\columnwidth]{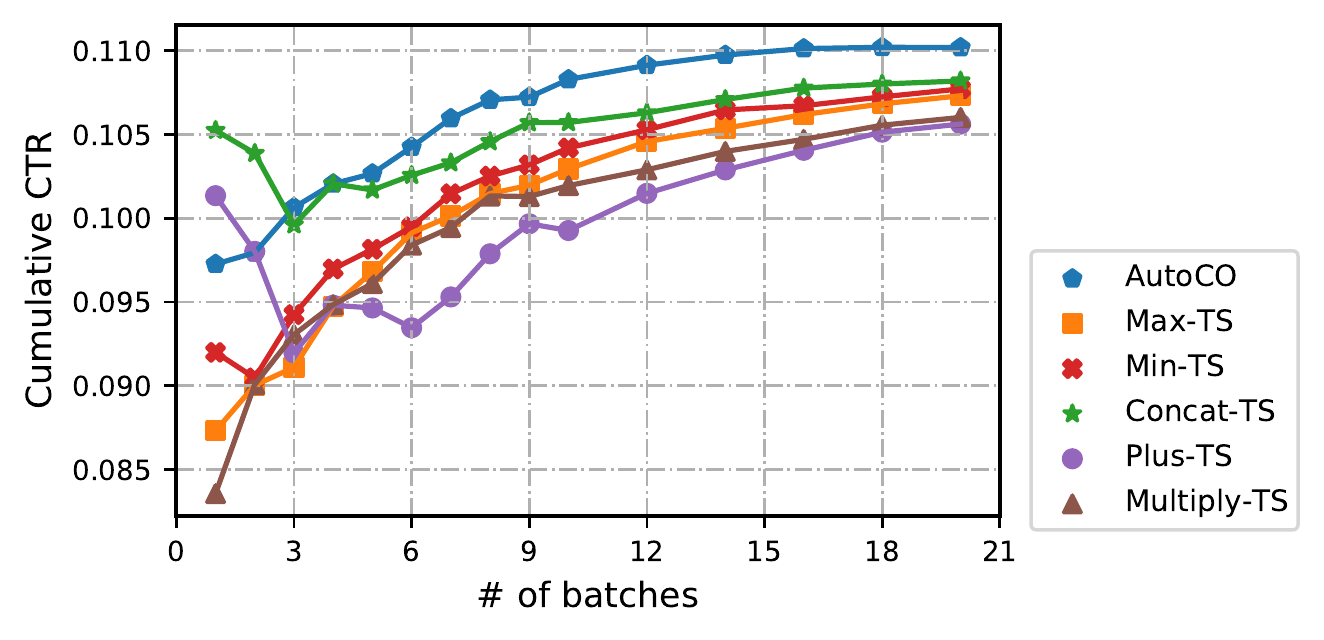}
	\caption{Comparisons of CTR on the different interaction functions.}
	\label{simulate_single_com}
\end{figure}

First, we compare the AutoCO with the fixed single operation in the search space of AutoCO. Results are shown in Fig. \ref{simulate_single_com}. All the algorithms are conducted with Thompson Sampling as the exploration method. Among the five operation functions, the concat operation has the best performance than the other four fixed operations while the AutoTS outperforms the concat interaction function. It can be inferred that better interaction functions for different feature fields are found out via the AutoML method. 



\begin{table}[ht]
	\caption{Comparisons of the overall CTR on different exploration methods. The relative improvements of the method exploration are marked in brackets.}
	\label{exploration_methods}
	\begin{threeparttable}
		\begin{tabular}{c|c|c}
			\toprule
			& FM & AutoCO$^*$ \\
			\midrule
			Greedy & 0.1030 & 0.1073\\
			Egreedy($\epsilon=0.1$) & 0.1024 (-0.58\%) & 0.1086 (+1.21\%) \\
			Thompson Sampling & 0.1055 (\textbf{+2.43\%}) & 0.1102 (\textbf{+2.70\%})\\
			\bottomrule
		\end{tabular}
		\begin{tablenotes}
			\item[*]\small AutoCO here denotes the interaction function search method for simplicity of understanding.  
		\end{tablenotes}
		
	\end{threeparttable}
\end{table}

Second, different exploration methods are tested to compare the efficiency with the Thompson sampling method utilized in this paper. For the FM model and AutoCO, experiments are run for five times to report the average performance in Table \ref{exploration_methods}. The variation based Thompson Sampling achieves excellent performances compared with Greedy and Egreedy. The trivial exploration method, Egreedy, would always recommend a predefined percentage ($\epsilon$) of creatives randomly although a large number of impressions are collected, causing a waste for displaying bad creatives. Instead, Thompson Sampling recommends the creative proportionally to the probability of being optimum and finally only the optimal creative would be displayed. Another point is that the variational inference for posterior approximation is well conducted in Thompson sampling and this method enables straightforward training in a variety of basic models.

In conclusion, the proposed AutoCO shows the superiority in the accuracy of the CTR predictor and efficiency for exploration on the synthetic data where better expressiveness is obtained through interaction function search.

\subsubsection{\textbf{Effectiveness of OAE}}
We conduct an ablation study on the simulated data to analyze the effect of the operation-aware embedding module. We first test the AutoCO and then remove the OAE module. The result is reported in Table~\ref{ablation_study}. The OAE module leads to a slight improvement in terms of CTR compared with the shared embeddings between different operations. This indicates the assistance of the independent embeddings. In the process of training, we also find that OAE contributes to more stable learning.

\begin{table}[h]
	\caption{Effectiveness of the operation-aware embedding (OAE) module. The experiments are run for 5 repetitions on the simulated data.}
	\begin{tabular}{c c}
		\toprule
		Framework & Overall CTR\\
		\midrule
		AutoCO w/o OAE & 0.1097 $\pm$ 0.0006\\
		AutoCO & \textbf{0.1102 $\pm$ 0.0014}\\
		\bottomrule
	\end{tabular}
	\label{ablation_study}
\end{table}

\begin{table*}[t]
	\caption{Comparisons with single interaction function on two public datasets. Results are relative to the cumulative regret to the Uniform algorithm. Three trials are conducted and we report the mean and standard error. Lower is better.} 
	\begin{tabular}{c|r|r|r|r|r|r|r}
		\toprule
		Dataset & AutoCO & Multiply-TS & Concat-TS & Plus-TS & Min-TS & MAX-TS & Uniform \\
		\midrule 
		Mushroom & \textbf{3.02$\pm$0.51} & 5.44$\pm$1.05 & 6.12$\pm$2.37 & 3.40$\pm$0.30 & 3.62$\pm$0.35 & 4.95$\pm$2.00 & 100.00$\pm$0.38\\
		Adult & \textbf{35.40$\pm$0.03} & 49.28$\pm$0.36 & 35.79$\pm$0.06 & 37.19$\pm$0.39 & 38.34$\pm$2.01 & 36.55$\pm$1.18 & 100.00$\pm$0.15\\
		\bottomrule
	\end{tabular}
	\label{NAS_operation}
\end{table*}

\begin{table*}[t]
	\caption{Comparisons for different exploration methods on public datasets Mushroom and Adult.} 
	\begin{tabular}{c|r|r|r|r|r|r}
		\toprule
		Dataset & AutoCO & AutoCO w/o TS & FM & FM-TS & Egreedy & Uniform \\
		\midrule 
		Mushroom & \textbf{3.02$\pm$0.51} & 5.58$\pm$0.51 & 7.70$\pm$1.44 & 5.64$\pm$1.70 & 63.38$\pm$0.47 & 100.00$\pm$0.38\\
		Adult  & \textbf{35.40$\pm$0.03} & 40.07$\pm$0.40 & 43.62$\pm$3.32 & 38.71$\pm$2.30 & 58.85$\pm$0.15 & 100.00$\pm$0.15\\
		\bottomrule
	\end{tabular}
	\label{NAS_explore_baseline}
\end{table*}

\subsection{Experiments on Public Datasets}
Empirical studies are also conducted on the two public datasets to verify the performances of our proposed AutoCO.
\subsubsection{\textbf{Datasets}}
\begin{itemize}
	\item \textbf{Mushroom}: The Mushroom dataset~\cite{schlimmer1981mushroom} contains 22 attributes and 8124 samples. The mushrooms are divided into two categories: safe and poisonous. We following the previous settings~\cite{blundell2015weight,riquelme2018deep} to formulate the bandit problem, where the algorithm needs to decide whether to eat the given mushroom. If eating a safe mushroom, a reward +5 will be received. Eating a poisonous mushroom obtains the reward +5 with a probability of 0.5 and the reward -35 otherwise. If the agent does not eat the mushroom, no reward is delivered. We sample 50,000 pieces of data from the original dataset.
	\item \textbf{Adult}: The Adult dataset~\cite{kohavi1996scaling} involves 14 attributes of personal information. The task is to determine if a person makes over \$50K a year or not. If the agent makes a right prediction, a reward 1 is delivered. Otherwise, no reward is given. This dataset contains 48842 samples. We pick the 9 categorical features of the mixed 14 original features to conduct our experiments.
\end{itemize}

\subsubsection{\textbf{Effectiveness of interaction function search}}
Compared with the single interaction function, the AutoCO has the lowest cumulative regret on the two public datasets, as shown in Table \ref{NAS_operation}. All the algorithms utilize the Thompson Sampling for exploration. The good performance of the AutoCO indicates that the complex interaction functions rather than the single operation enhance the performance of modeling relationships between the different attributes.

\subsubsection{\textbf{Efficiency of exploration methods}}
To verify the efficiency of Thompson sampling via variational inference, we conduct experiments to compare it with greedy strategy. Experiment results are shown in Table \ref{NAS_explore_baseline}. We can conclude that Thompson sampling via variational inference can indeed increase the performance.

\subsection{Online Experiment}
Now, we examine the performance of our proposed algorithm on the online system.

\begin{figure}[t]
	\includegraphics[width=\columnwidth]{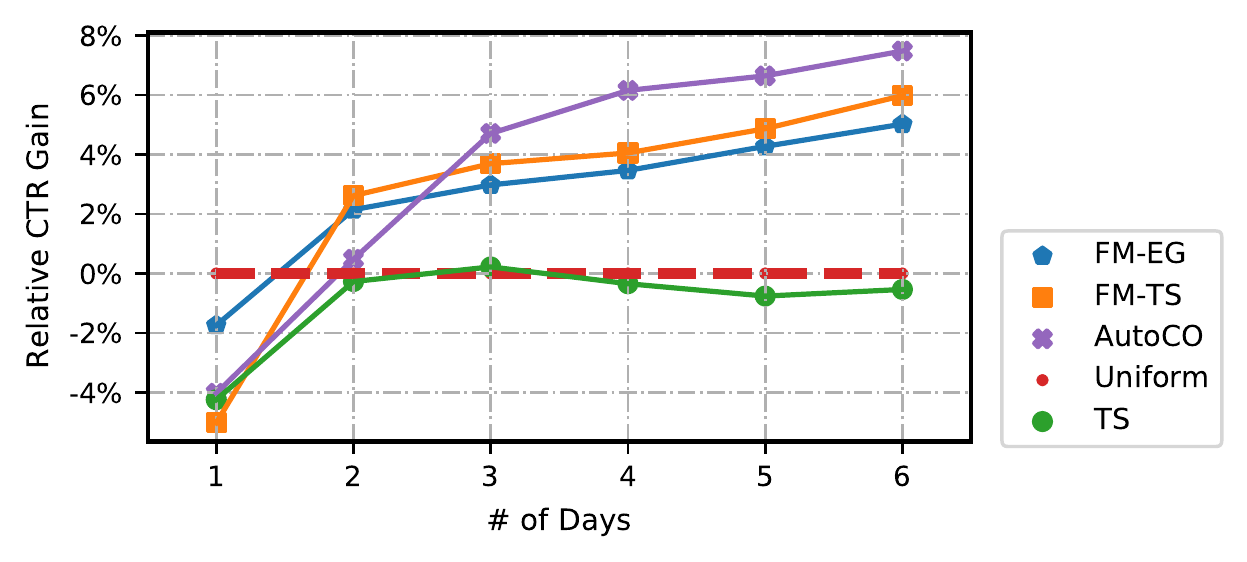}
	\caption{Cumulative CTR of online experiments. We report the relative improvements compared with the Uniform algorithm. There are more than 150,000 impressions per algorithm per day, covering more than 290,000 creatives.}
	\label{online_experiments}
\end{figure}

\subsubsection{\textbf{Experimental Design}}
We exploit AutoCO for the recommendation of the composited creatives in an online advertisement platform. The composited elements include the template, picture size, text font, background blur, and background color. Our experiments are conducted on one display advertising position of a famous e-commerce company. We generate more than 290,000 creatives satisfying visual constraints covering 1,000 popular products. In addition to the element feature mentioned before, we utilize the category feature of products as well as the contextual information of product images. The feature embeddings of images are extracted from ResNet.

We performed the experiments for six consecutive days. Traffic for this experiment was randomly and equally separated to the algorithms. The experiments were conducted on non-holidays to avoid fluctuations in traffic. Each algorithm got more than 150,000 impressions per day. The model was updated every hour using the data collection from the beginning to the latest hour.

Baseline algorithms: (1) \textit{Uniform}: The composited creatives are delivered to the platform randomly; (2) \textit{Context-free Thompson Sampling}: The distribution of expected reward is assumed as a Beta distribution. The initialized $\alpha=1$ and $\beta=50$; (3) \textit{FM-Egreedy}: FM is the CTR predictor and $\epsilon=0.2$; (4) \textit{FM-TS}: Apply the proposed Thompson method for exploration in the FM model. In our offline experiments, we found that the interaction functions do not change frequently after several epochs, which is also declared in~\cite{yao2019efficient}. In order to reduce the burden of online updating, we do not frequently search the interaction functions.

\subsubsection{\textbf{Result}}
The six days online experiment shows the superiority of our approach AutoCO, as shown in Fig.~\ref{online_experiments}. The context-free TS algorithm almost has no improvements compared with the Uniform baseline, which exactly demonstrates the sparsity of feedback. On average, the impression for each creative is less than one per day. The contextual algorithms have better performances where the interaction between different features are paid attention to capture the commonality between different creatives.

On the online A/B test, although the AutoCO is merely comparable in the first two days, the significant improvement is achieved on day 3. In the following consecutive three days, the AutoCO steadily improve the performance. This indicates the quick convergence of the proposed method which is essential for online advertisement. Compared with the FM-based models, our approach still shows competitive performances. It can be inferred that the search of interaction functions help find out more appropriate interaction functions to improve the expressiveness. 

\section{Conclusion}
In this paper, we propose an automated creative optimization framework to model complex interaction between creative elements and to strike a balance between exploration and exploitation. For modeling complex interaction, we suggest an one-shot search algorithm for searching more effective interaction function in the customized function space efficiently. For balancing between exploitation and exploration, we develop the stochastic variational inference for posterior approximation based on the reparameterization trick, and then apply Thompson Sampling for effective exploration of potentially better creatives. We conduct both offline and online experiments, showing that our proposed approach performs better than the competing baselines and verifying the effectiveness of the proposed algorithms.

\begin{acks}
The work was supported by Alibaba Group through Alibaba Innovative Research Program. Defu Lian is supported by grants from the National Natural Science Foundation of China (Grant No. 61976198, 62022077),  the National Key R\&D Program of China under Grant No. 2020AAA0103800 and the Fundamental Research Funds for the Central Universities. Kai Zheng is supported by NSFC (No. 61972069, 61836007, 61832017) and Sichuan Science and Technology Program under Grant 2020JDTD0007.
\end{acks}

\bibliographystyle{ACM-Reference-Format}
\bibliography{myreference}

\appendix

\end{document}